\pdfoutput=1

\documentclass{article}

\PassOptionsToPackage{numbers}{natbib}

\usepackage[final]{nips_2017}

\usepackage[utf8x]{inputenc} 
\usepackage[T1]{fontenc}    
\usepackage{hyperref}       
\usepackage{url}            
\usepackage{booktabs}       
\usepackage{amsfonts}       
\usepackage{nicefrac}       
\usepackage{microtype}      
\usepackage{graphicx}
\usepackage{subfig}
\usepackage{xfrac}
\graphicspath{ {diagrams/} }
\usepackage[autostyle]{csquotes}
\usepackage{todonotes} 

\title{Neural Translation of Musical Style}

\author{
  Iman Malik \\
  Department of Computer Science\\
  University of Bristol\\
  Bristol, U.K \\
  \texttt{im13557@my.bristol.ac.uk} \\
  \And
  Carl Henrik Ek \\
  Department of Computer Science\\
  University of Bristol\\
  Bristol, U.K \\
  \texttt{carlhenrik.ek@bristol.ac.uk} \\
}

\begin{document}

\maketitle

\begin{abstract}
  Music is an expressive form of communication often used to convey emotion in scenarios where ``words are not enough''. Part of this information lies in the musical composition where well-defined language exists. However, a significant amount of information is added during a performance as the musician interprets the composition. The performer injects expressiveness into the written score through variations of different musical properties such as dynamics and tempo. In this paper, we describe a model that can learn to perform sheet music. Our research concludes that the generated performances are indistinguishable from a human performance, thereby passing a test in the spirit of a ``musical Turing test''.

\end{abstract}

\section{Introduction}\label{sec:introduction}
Music is mysterious. Anthropologists have shown that every record of human culture has some aspect of music involved \cite{Morley2014}. However the exact evolutionary role of music is shrouded in mystery. Scholars theorise and state that music must have emerged as an evolutionary aid \cite{Schulkin2014,Huron2008}. One theory proposes that music may have arisen from mothers putting their children to sleep \cite{Falk2004}. Some propose that the function of music was to provide social cement for group action \cite{Schulkin2014, Mithen2006, Kniffin2016}. War songs, national anthems, and lullabies are all examples of this.

Music is fundamentally a sequence of notes. A composer constructs long sequences of notes which are then performed through an instrument to produce music. Often these songs possess the ability to convey an emotional and psychological experience for the listener \cite{Perlovsky2010,Lundqvist2008}. Two important aspects of music are the composition and the performance \cite{deMantaras}. The composition focuses on the notes which define the musical score. Over centuries humans have developed different ways of transcribing musical compositions usually referred to as sheet music \cite{Schulkin2014a}.

However, when music is performed from sheet music, it needs to be interpreted. The ambiguity during interpretation results in a variety of different realisations of the same sheet description. In abstract terms, this means that the mapping between the sheet notation and the performed music is not a bijection. A classic example of this are cover songs, \citet[][p. 1]{ellis07_ident_cover_songs_chrom_featur} stated that \enquote{\it Indeed, in pop music, the main purpose of recording a cover version is often to investigate a radically different interpretation of a song}. This characteristic is what makes automatic music synthesis challenging, as we are looking to discover a multi-modal mapping. Musical style is challenging to parameterise and contradictory to the idea of a cover song, as it is often attributed to all aspects of the song \cite{DBLP:conf/ismir/MayerR11}. With music being one of the pioneering digital domains with over 43 million songs licensed digitally in 2016 \cite{digitalmusic16}, there exists a wealth of musical data to learn from. This leads the central question of this paper, {\it is it possible to leverage data and learn how to automatically synthesise musical performances that are indistinguishable from a human performance?} Specifically, we postulate that a significant portion of the style injected by a musician comes from \textbf{dynamical aspects}. To that end, we aim to learn to inject the note \textbf{velocities} from data only containing the note \textbf{pitches} over time.

The remainder of the paper is structured as follows. In Section~\ref{sec:methodology}, we will describe our model and how it relates to previous work. We will then proceed to described the experimental setting and the results in Section~\ref{sec:experiments} and Section~\ref{sec:results} respectively. We will then conclude the paper and provide some directions for future work in Section~\ref{sec:conclusions}.

\section{Related Work and Methodology}\label{sec:methodology}
Music plays an important role in many peoples' lives. Thus it is not surprising that several works focus on the complicated problem of music synthesis. Several attempts have been made at generating musical compositions. One of the earliest generative models, ``CONCERT'', was architected to compose simple melodies \cite{Schulze2011}. However, the limitations of CONCERT were that it could not capture the global structure of music. The generated music was said to lack ``global coherence''. This is problematic as music has long-range dependencies. Based on the CONCERT model, \citet{Eck2002} tackled this problem by building a model that could learn longer-range dependencies.

These models can be labelled as ``compositional'' models. There have been several attempts to train performance models which focus on capturing the performers' touch through features such as dynamics, tempo, and so on. One of the earlier performance models, ``Director Musices'', which was a rule-based model incorporating rules inferred from theoretical and experimental knowledge \cite{Friberg}. However such rule-based models cannot cannot capture the large variations in performances as they cannot learn new rules. Such approaches were then superseded by rule learning approaches \cite{Widmer2003, stanislas}.

Our aim is to predict the note velocities from a sequence of notes, which implies that we are learning in a regression scenario. In recent years, neural networks have re-entered the forefront of machine learning research. For tasks where data is abundant, feedforward neural networks are pushing the boundaries of the tasks that machine learning can solve. These types of networks are very general and make no assumption on the structure of the data. Music is highly dynamic, therefore we must ensure that the model accommodates for this property. Recurrent Neural Networks (RNN) \cite{lipton15_critic_review_recur_neural_networ_sequen_learn} are designed to capture dynamic structures by retaining a ``memory'' of previous patterns. A recent approach successfully used RNNs to capture the style of different pianists \cite{stanislas}. However not much research has been done on different genres of music.

\begin{figure}[b]
\centering
\includegraphics[width=0.6\textwidth]{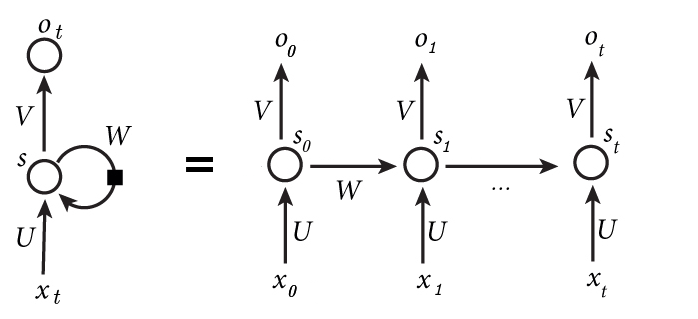}
\caption{An unfolded RNN.}
\label{unfolded}
\end{figure}

We denote the RNN's input and output as $x_{t}$ and $o_{t}$ respectively as seen in \autoref{unfolded}. The RNN has three main parameters $U, V, W$. The weights $U$ and $V$ correspond to the input $x_{t}$ and output $o_{t}$ respectively. The recurrent weight $W$ determines how much of the previous state will be introduced into the RNN's immediate computation, and is shared across all time-steps.

As mentioned above, RNNs can be effective when processing sequences. However, the RNN suffers from the vanishing gradients problem. This would be problematic when long-term dependencies or context needs to be captured in a musical piece. This motivates a special type of RNN called the Long Short-Term Memory Network (LSTM) which was specifically designed to avoid such issues \cite{Hochreiter1997}.

\begin{figure}
\includegraphics[width=0.4\textwidth]{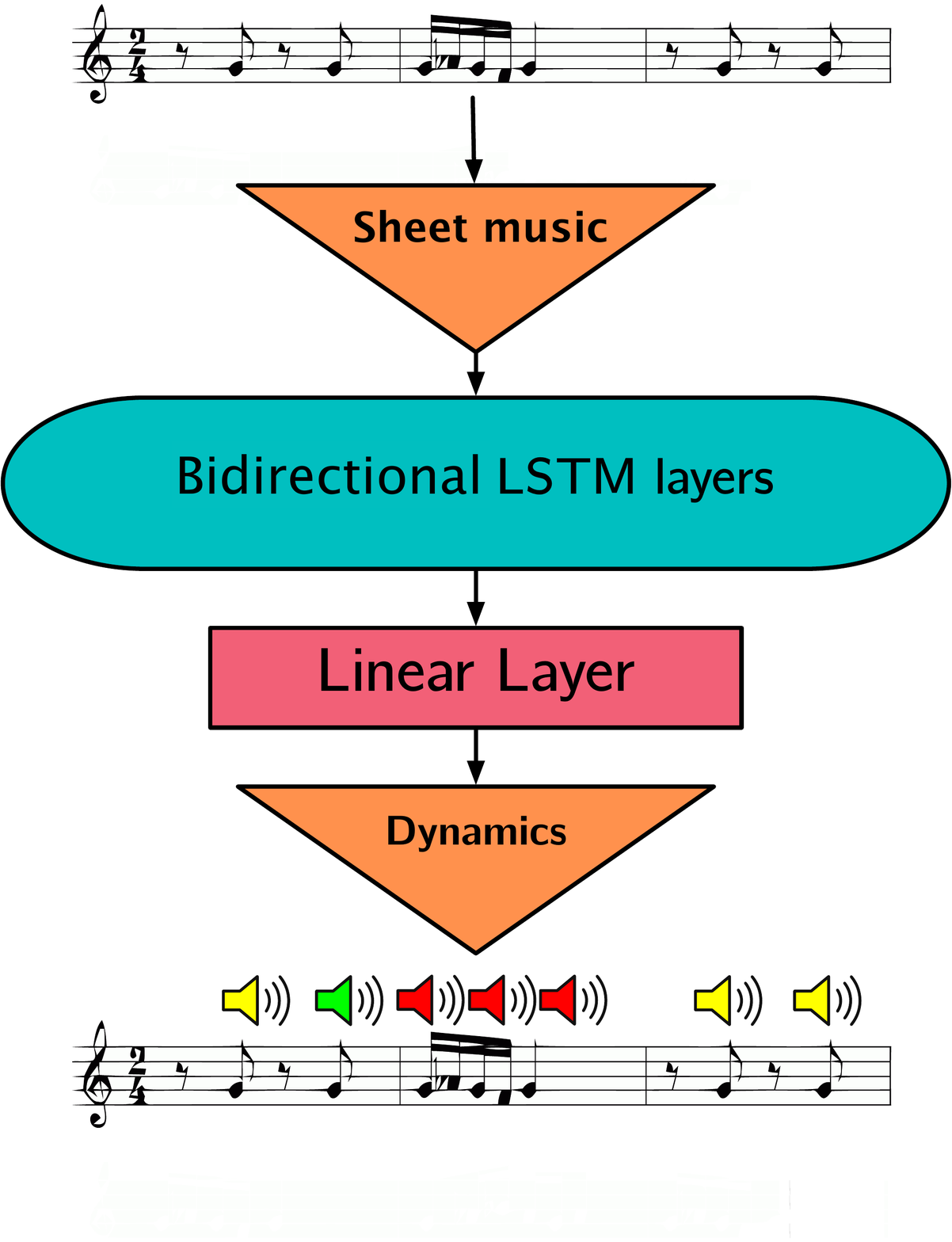}
\centering
\caption{GenreNet}
\label{fig:genrenet}
\end{figure}

With the motivation mentioned above, the intuition behind the initial design of the network can be explained. To learn style, one needs to first focus on a subset of the problem. Musical styles can be categorised by genre. We describe the architecture of GenreNet. GenreNet predicts the dynamics of a musical input such as sheet music. The model consists of two main layers as seen in \autoref{fig:genrenet}: the bidirectional LSTM layers and the linear layer.

\textbf{The Bidirectional LSTM layers}:
The bidirectional architectural choice is based on the real task of reading sheet music. Humans can use their sight to skim across sheet music and glance at upcoming notes in the score. They can use this visual ``look ahead'' to modify their performance. This would be analogous to using a bidirectional LSTm layer give us this foresight.

\textbf{The Linear Layer}:
To scale the output to represent a larger range of values, a linear layer can be used. A linear layer performs a linear transformation on its input. This transformation is called the identity activation function where $z$ is the weighted sum of its inputs.

\begin{equation}
f(z) = z = \mathbf{w}^Tx
\label{eq:indentity_activation_function}
\end{equation}

\subsection{StyleNet}
GenreNet is limited to learning the dynamics for a specific genre. However as stated in the introduction, the goal of this research investigates whether it is possible for a machine to learn to perform music like a human. Humans can play music in a variety of styles. This motivates the design of StyleNet, the rendition model.

In the field of computer vision, \citet{Bromley1993} introduced a neural network architecture called the Siamese Neural network. This architecture consists of identical subnetworks which share parameters. The purpose of this architecture is to learn the similar feature shared between two inputs.

\begin{figure}[h]
  \centering
\subfloat[]{\includegraphics[width=0.45\textwidth]{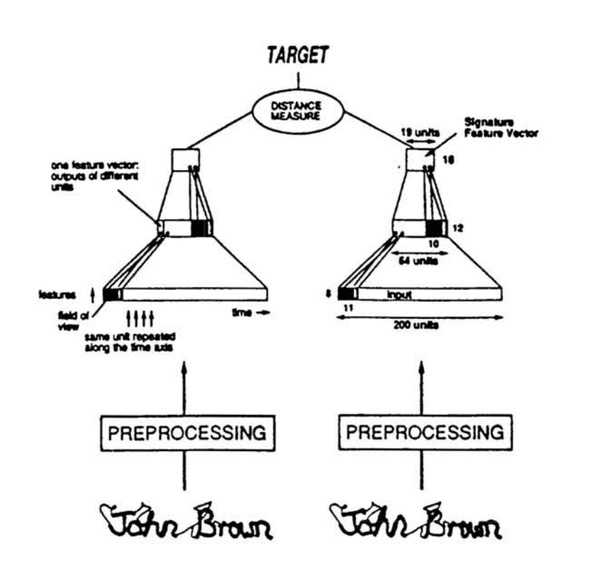}\label{siamese}}
  \hfill
  \centering
\subfloat[]{\includegraphics[width=0.45\textwidth]{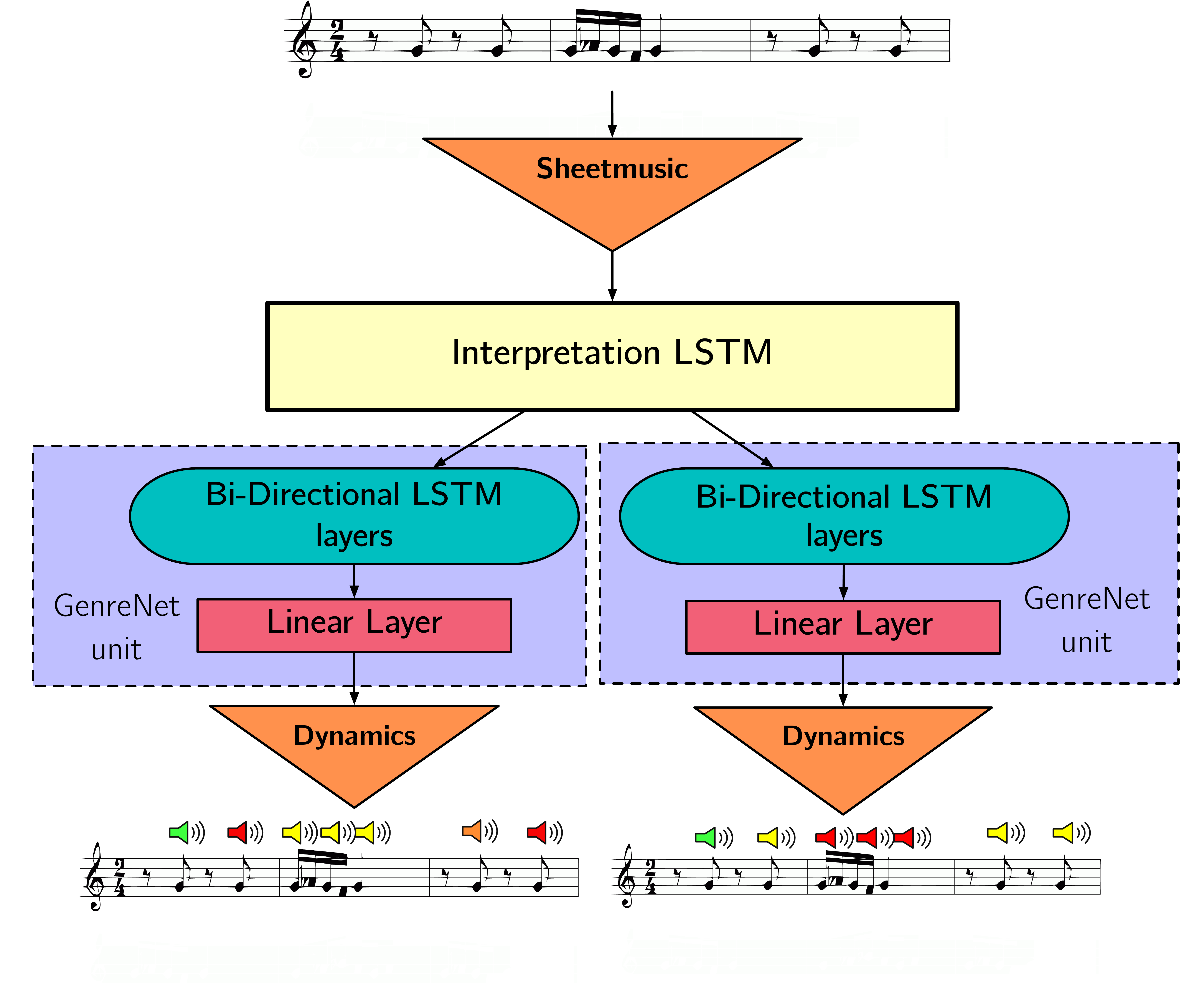}\label{fig:stylenet}}
  \centering
\caption{ (a) Siamese Neural Network Architecture \protect \cite{Bromley1993}. (b) StyleNet.}
\end{figure}

However, in this case, the similar feature is known. This feature is the sheet music. The task at hand is to produce different outputs for the sheet music. The StyleNet architecture has two main components as seen in Figure\autoref{fig:stylenet}: the interpretation layer and the GenreNet unit.

\textbf{Interpretation Layer}:
This is the shared layer across GenreNet units. The interpretation layer converts the musical input into its own representation of the sheet music. As this layer is shared, the number of parameters the network needs to learn are reduced. This ultimately leads to needing less data to train our model on which is always advantageous.

\textbf{GenreNet Unit}:
These subnetworks are attached to the interpretation layer. Each GenreNet unit allows the model to learn a specific style.

\section{Experiments}\label{sec:experiments}
Now that the StyleNet architecture has been designed, the training data needs to be obtained. The goal is to create a dataset from which StyleNet can learn Classical and Jazz style. We present the Piano dataset. The dataset contains Piano MIDI files within the Classical and Jazz genre. All MIDI files are in \sfrac{4}{4} time and format 0. Both genres have 349 MIDI files which creates a total of 698. The dataset will be available as complementary material.

\textbf{MIDI files}:
We choose the MIDI file format because it already contains musical metadata such as note velocities unlike WAV. There are numerous MIDI files available on the internet.

\textbf{Isolating Genre}:
Since we are working within the limitations of the MIDI format, most human-performed recordings are of piano and drum MIDI controllers. The piano plays a dominant role in both Jazz and Classical, and thus the focus will be on these two genres.

\textbf{Isolating Piano}:
Across Jazz and Classical MIDI files, there are several instruments. We decide to focus on the dynamics of the piano.

\begin{figure}
  \centering
\subfloat[All downloaded MIDI files.]{\includegraphics[width=0.45\textwidth]{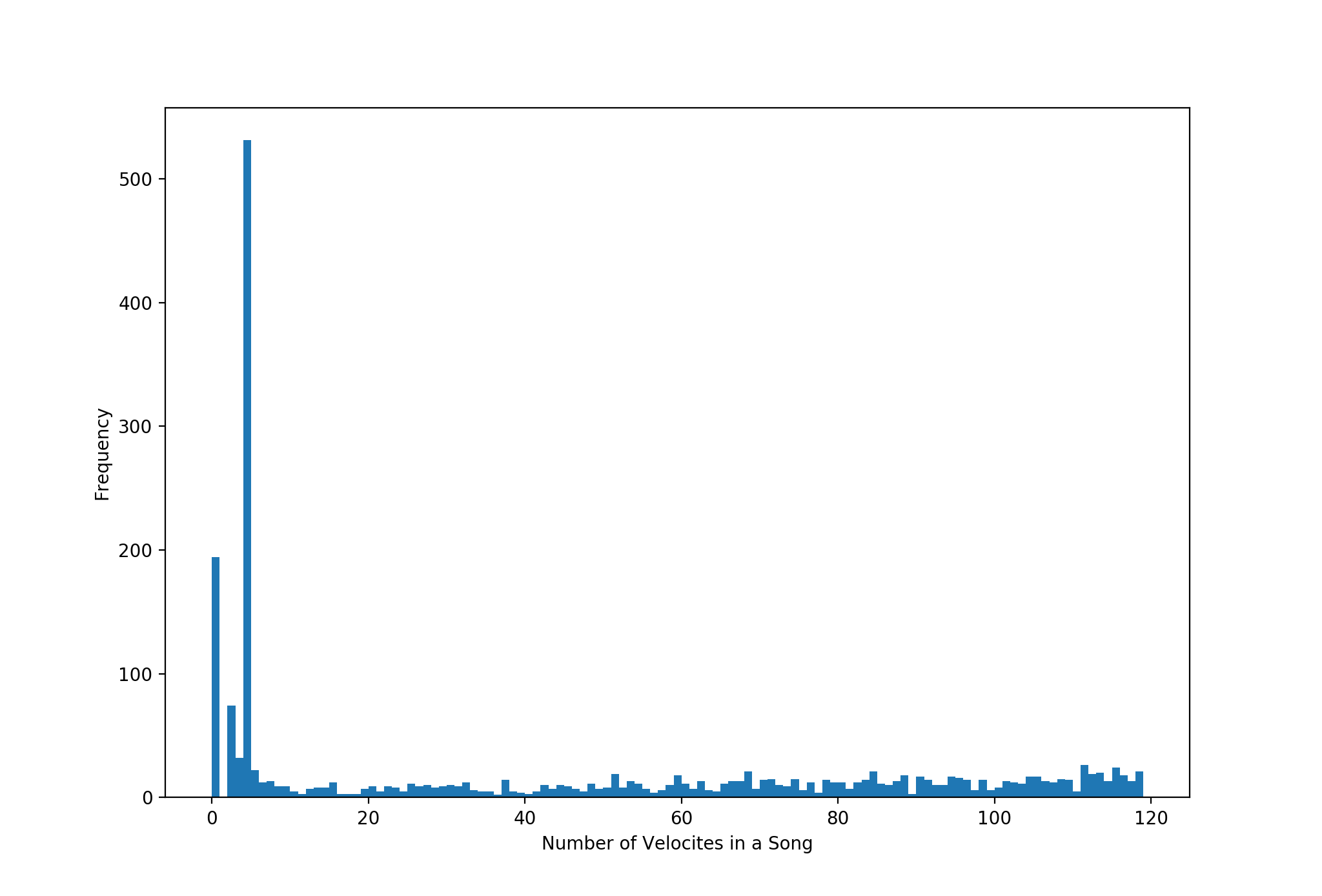}\label{downloaded_midi}}
  \hfill
\subfloat[Performance MIDI files.]{\includegraphics[width=0.45\textwidth]{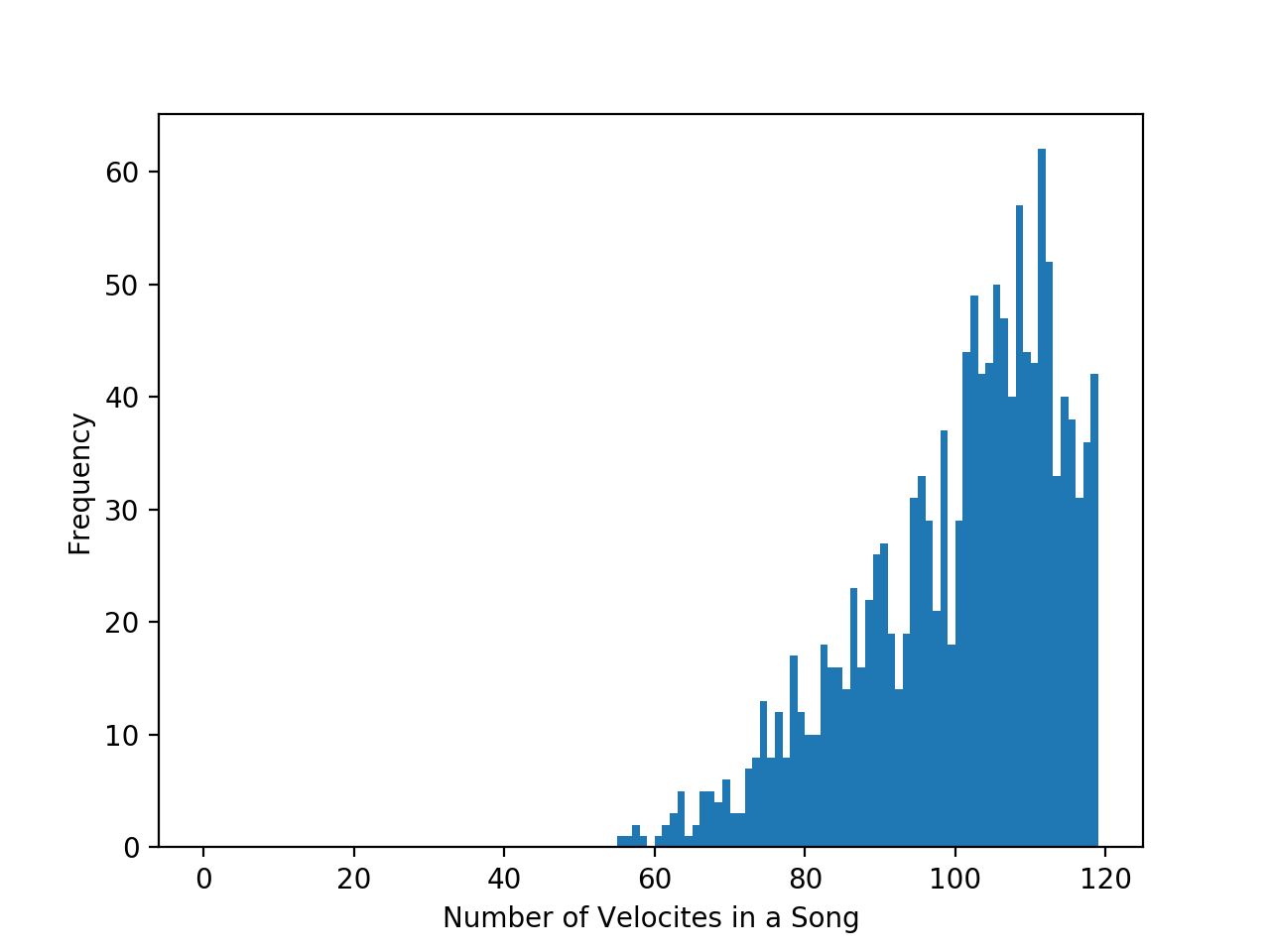}\label{performance_midi}}
\caption{Histograms of velocity range across MIDI files.}
\end{figure}

\textbf{Capturing Velocity}:
Many software-generated tracks only contain one global velocity. This can be seen in Figure\autoref{downloaded_midi}. This is noticeable in the large quantity of MIDI files with $10$ or less different velocities. Using a baseline from live performance MIDI files \cite{yamaha}, a minimum threshold of at least 20 different velocities was chosen for the dataset.

\begin{figure}
  \centering
\subfloat[]{\includegraphics[width=0.45\textwidth]{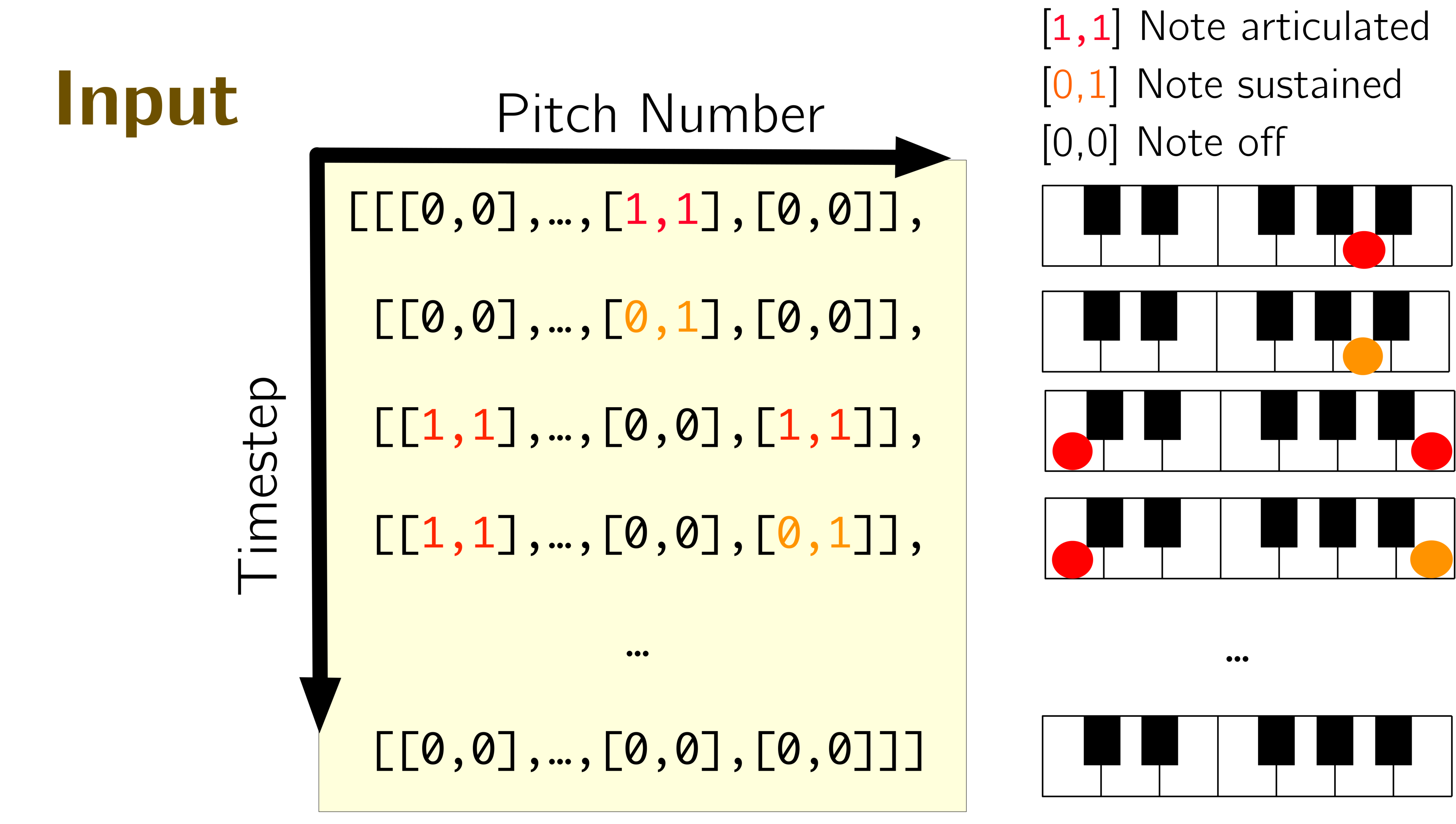}\label{input}}
  \hfill
  \centering
\subfloat[]{\includegraphics[width=0.45\textwidth]{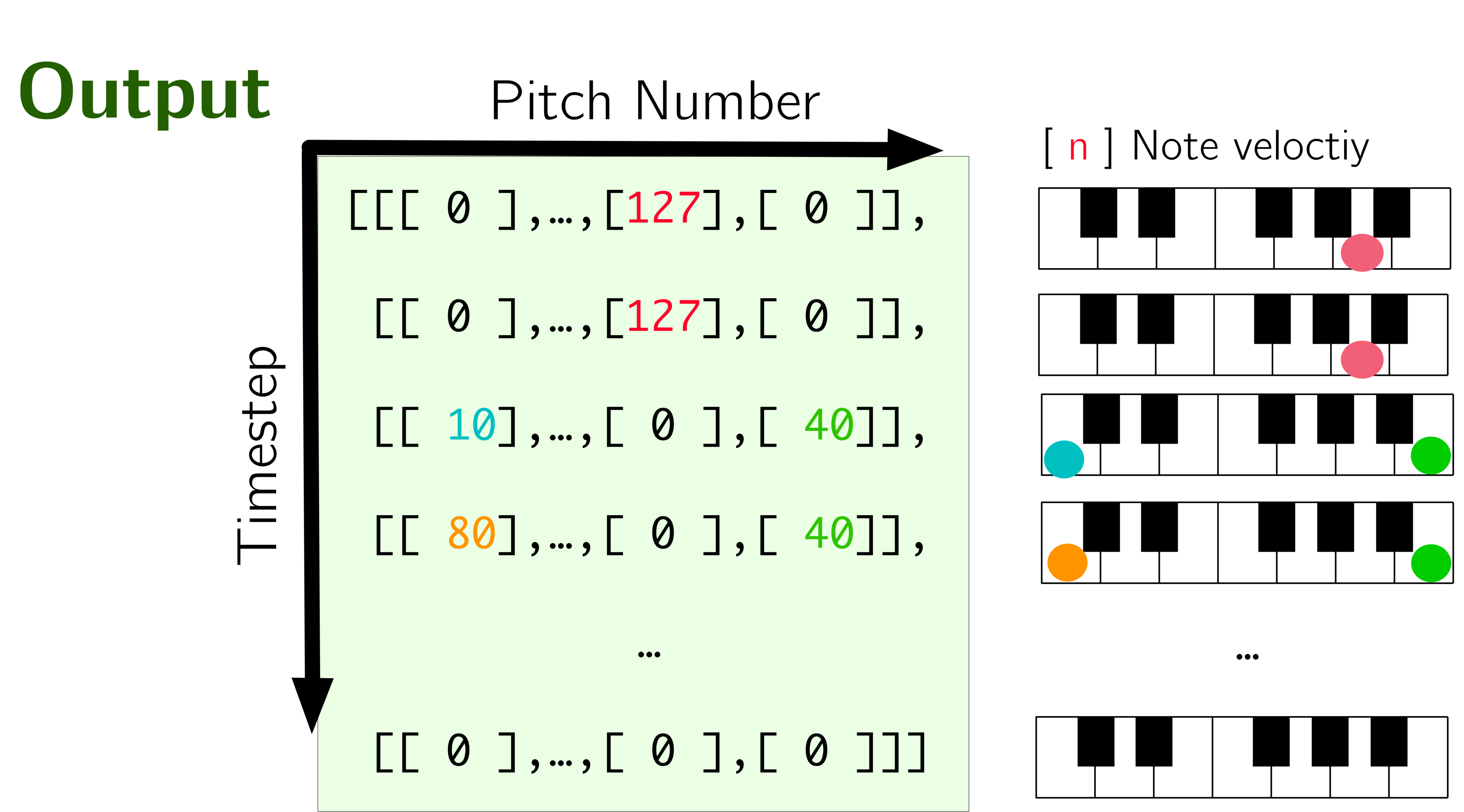}\label{output}}
  \centering
\caption{ Data representation matrix.}
\end{figure}

\textbf{Time Signature }:
Time is continuous. Unfortunately, we need to discretise/quantise our notes in order to represent them in a way our model can process them. To maximise the amount of data captured across the dataset, only songs with the same time signature were kept. \sfrac{4}{4} is most common and thus was chosen.

\textbf{Input Representation}:
Isolating important features is the first step to designing an input format. The model needs to know what notes are being played at a given time-step. A note can have three states: note is on, note is off, or note is sustained from the previous time-step.

Using a binary vector, ``note on'' is encoded as $[1,1]$, ``note sustained'' as $[0, 1]$ and ``note off'' as $[0, 0]$. The first bit represents whether the note was played in that time-step or not and the second represents if the note was held or not.

Next, the note pitch needs to be encoded. At one time-step, any possible note pitch could be played. Recalling that MIDI encodes pitch as a number in the range $[0,127]$, a matrix with the first dimension representing MIDI pitch number is created. The second dimension represents a quantised time-step or a \sfrac{1}{16} note.

\textbf{Output Representation}:
Similar to input matrix above, the columns of our matrix represents pitch and the rows represent time-step. The velocities of the notes are encoded into the matrix. The velocities are preprocessed, and are divided by the max velocity $127$ so the network does not have learn the scale itself. This means all the velocities are between $0$ and $1$.


Training neural networks requires a strong understanding of their underlying theory \cite{Pascanu2012}. The goal of StyleNet is to learn Jazz and Classical styles. We will describe the setup and the series of experiments done to justify the final hyperparameters for StyleNet. Our training and validation are set to be 95\% and 5\% respectively.

\textbf{Model}:
The input interpretation layer is set to be 176 nodes wide and only one layer deep. There are two GenreNet units: one for Jazz and one for Classical. Each GenreNet is three layers deep.

\textbf{Loss function}:
StyleNet outputs a velocity matrix for each genre through its GenreNet unit. This is a regression learning problem. A metric to measure the performance of the model would be the mean squared error (MSE) between the true and predicted velocity matrix. $X$ represents the music input and true velocity output vector pairs, $X = \{(x_{1},y_{1})...(x_{N},y_{N})\}$, $N$ is the number of time-steps in a song, and the $h$ is the network's prediciton and is parameterised by $\theta =\{W, b\}$

\begin{equation}
E(X) = \frac{1}{N}\sum_{i=1}^{N}(h_{\theta}(x_{i})-y_{i})^{2}
\label{MSE}
\end{equation}

\textbf{Truncated Backpropagation Through Time}:
Backpropagation is truncated to 200 time-steps to reduce training time. This limits our model to learn dependencies within a 200 time-step window. However, this improved training time significantly. Convergence time was reduced from 36 hours to around 12 hours with truncation.

\textbf{Dropout}:
A dropout of $p = [0.5, 0.8]$ was experimented with using a learning rate of $10^{-3}$. However, the model would underfit on a dropout of 0.5. Thus a dropout value of $0.8$ is chosen.

\textbf{Gradient Explosion}:
 LSTM networks are vulnerable to having their gradients explode during training. We clip the gradients by norm \cite{Pascanu2012b}. This method introduces an additional hyperparameter called $g$. When the norm of a calculated gradients is greater than $g$, then the gradient is scaled relative to $g$. This parameter is set to 10.

\textbf{Final Model}:
Now the setup and results for the final model as can be listed. The StyleNet was successfully trained on alternating batches of Jazz and Classical music using the Adam optimiser on a Nvidia GTX 1080 Ti.  A dropout of $p = 0.8$ was applied, and gradients were clipped by norm where $g = 10$ with a learning rate of $10^{-3}$. The model was training for a total of 160 epochs. The final and validation loss was $7.0\times10^{-4}$ and $1.1\times10^{-3}$ respectively.

\begin{figure}[h]
\includegraphics[width=0.95\textwidth]{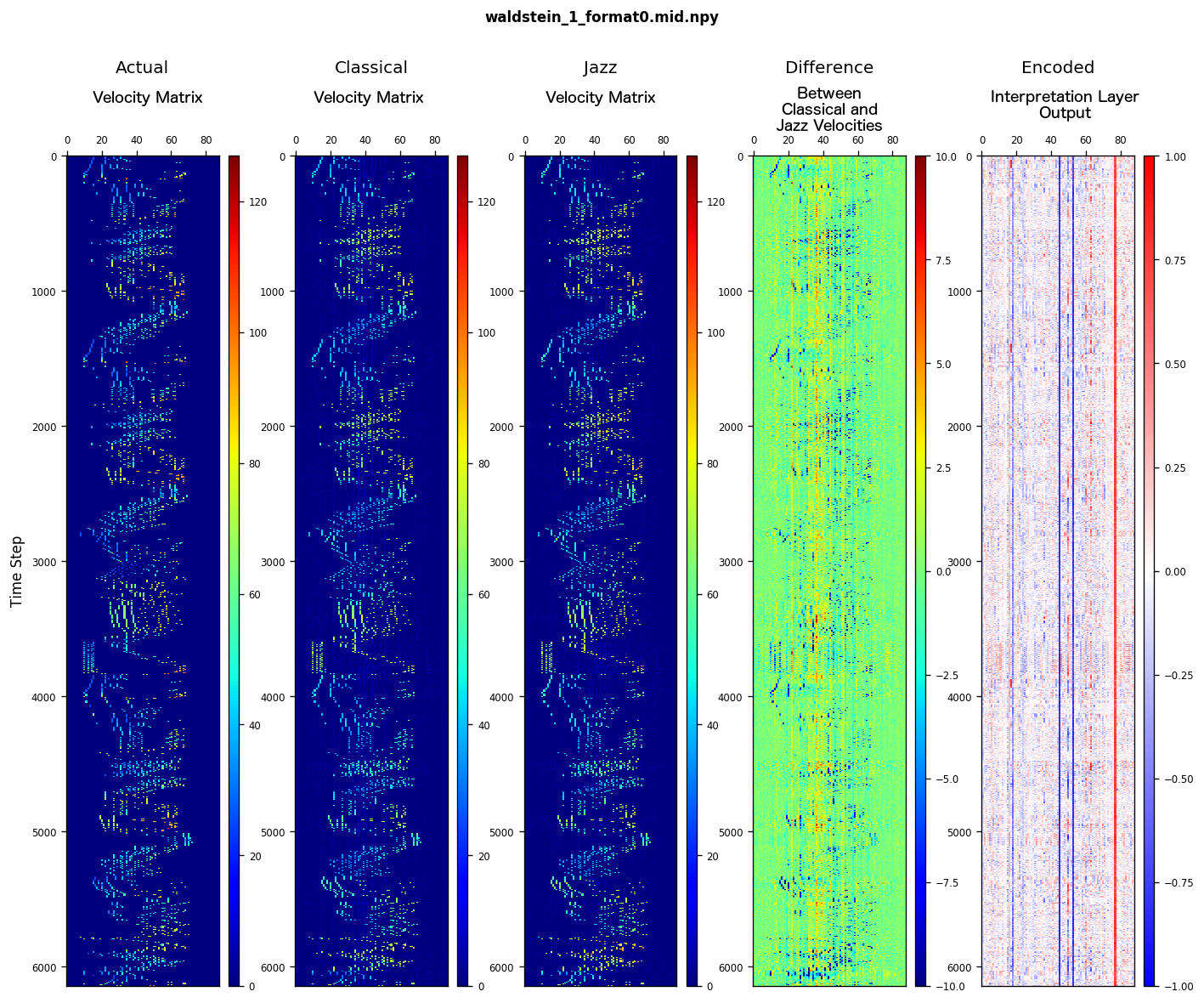}
\centering
\caption{Training snapshot of StyleNet's predictions for waldstein\textunderscore 1\textunderscore format0.mid.}
\label{snapshot}
\end{figure}

\section{Results}\label{sec:results}
How does one evaluate a musical performance? Music only holds meaning through the confirmation of a human. The decreasing loss shows us that the model is trying to understand the problem numerically. However what one wants is to minimise the ``perceptual'' loss. Thus it can be quite challenging when trying to evaluating a model in the field of music.

As mentioned in the introduction, the primary objective is to investigate whether a machine can perform sheet music like a human. Alan Turing's Turing test will be taken as inspiration for the evaluation \cite{Alan1950}.

Three experiments are conducted. ``Identify the Human'' is a musical Turing test. This was performed twice. First on short and then on long audio clips. The other experiment, ``Identify the Style'' investigates whether the model has learned style. The validation set was used to generate performances for the experiment.

\textbf{''Identify the Human'' Test}:
The ``Identify the Human'' survey was set up in two parts with 9 questions each. For each question, participants are shown two 10 second clips of the same performance. One performance is generated and the other is an actual human performance. Participants need to identify the human performance. The ordering of the generated and human tracks was randomised to reduce bias towards a particular answer.

An average of 53\% from the participant pool could highlight the human performance. There is no known benchmark for this problem. Thus a baseline is a random guess. This reveals that on average, 3\% from the participant pool could perform better than random guessing. This is a surprisingly low number and concludes that the model passed the Turing test.

\textbf{''Identify the Style'' Test}:
This leads the next investigation into the model's ability to play sheet music in a specific style. The ``Classical or Jazz'' survey was set up in two parts with 9 questions each. Sheet music for a single performance is generated in a Classical and Jazz style. These two stylised tracks are shown to the participants. The task at hand for participants is to correctly identify the style being asked for.

An average of 47.5\% respondents selected the correct style. Similar to the previous test, the baseline of this test is randomly guessing between both answers. The analysis of this number shows that the structure of the Style model is not sufficient to separate the characteristics between the two styles. We believe that this could be the result of several different factors, for one, we do not have examples of the same sheet interpreted in both styles. Such data would encourage the style split at the interpretation layer in the model. Furthermore, style is something that is ``added'' to composition which might be challenging to capture with this sequential structure.

\textbf{Final ``Identify the Human'' Test}
As mentioned earlier, some participants mentioned that 10 seconds is not long enough to determine the human performance. It can be hard to assess a short clip without its surrounding music context. Thus a more valid Turing test would be to assess the model on a complete performance. This motivates this final Turing test.

\begin{figure}[h]
\includegraphics[width=0.4\textwidth]{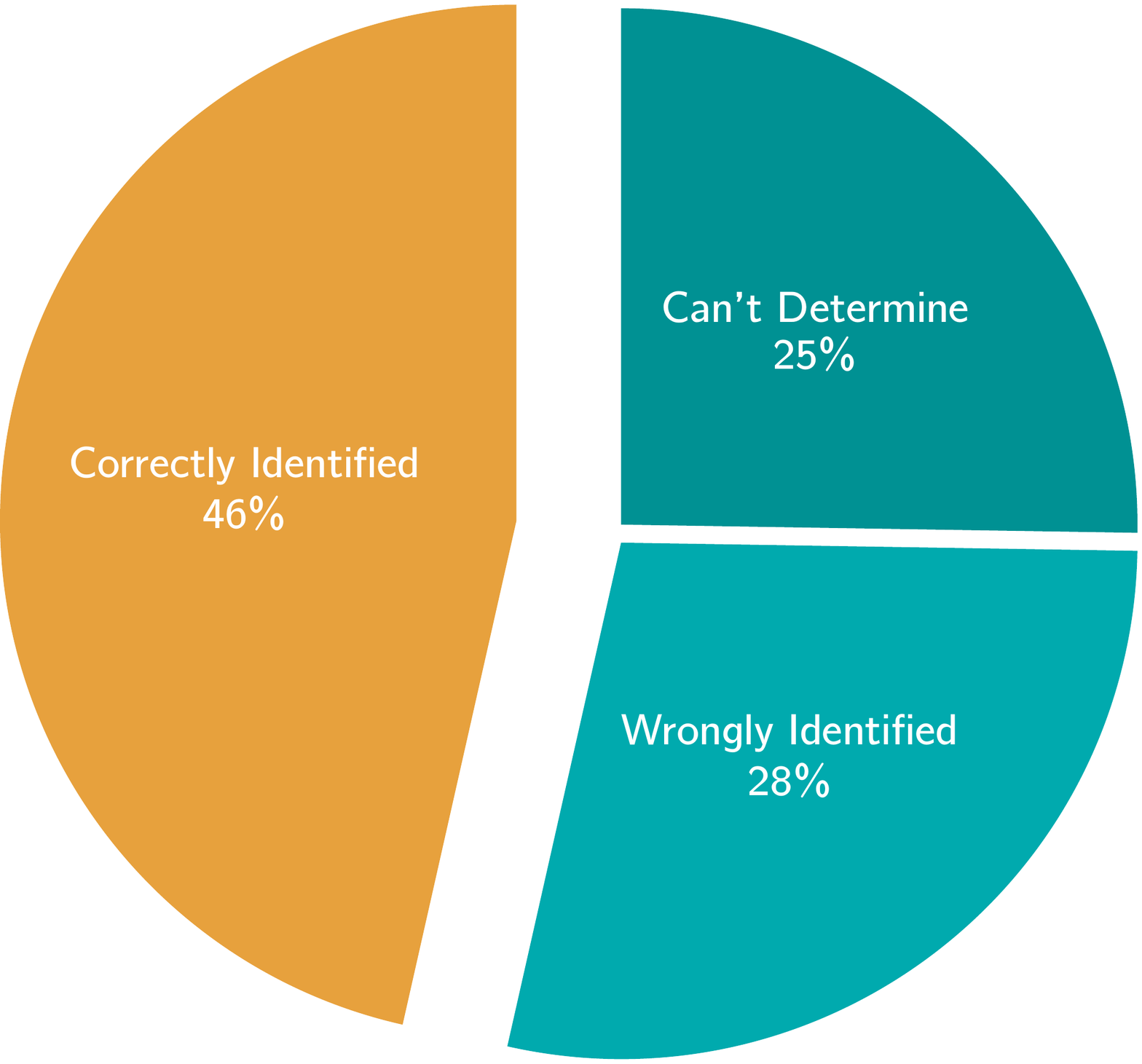}
\centering
\caption{Final ``Identify the Human'' survey results. }
\label{final_survey}
\end{figure}

The experiment set-up was identical to the ``Identify the Human''  test for short audio clips, but the only difference is that participants had to answer one question featuring an extended performance. The song used for this experiment was ``chpn-p25.mid'' which is a 2:30 Classical piece called ``Etudes Op.25'' by Fr\'{e}d\'{e}ric Chopin.

The survey was completed by 99 people. \autoref{final_survey} shows that only 46\% participants could identify the human. This shows that humans are not capable of differentiating between synthetic and real music. This concluded that StyleNet has successfully passed the Turing Test and can generate performances that are indistinguishable from that of a human.

\subsection{Summary of Results}
To summarise, three experiment have been successfully carried out on the trained StyleNet model. The first musical Turing test experiment, ``Identify the Human'', was performed on short audio clips. The results of this experiment concluded that participants could not tell the difference between short generated and real performances. The second experiment ``Identify the Style'' concluded that participants cannot correctly identify the style of the generated performances. This result leads to say that the model cannot generate noticeably stylised performances. The last experiment ``Identify the Human'' concluded that participants could not tell the different between the two extended performances. The results of this experiment strengthen our initial findings.

\section{Conclusion}\label{sec:conclusions}
In this paper we have presented a model that is capable of creating natural sounding performances which are indistinguishable from a human performance. Our style model is based on a LSTM network. We also experimented with separately modelling style from content in order to translate music between different genres. Our results shows that this approach was not suitable for the task and additional work is required. We have also created the Piano dataset which is publicly available to allow for further research in this exciting area.

In our future work, we want to focus on learning decompositions of music which separates style from content. The StyleNet model proposed in this paper was not sufficient for this task. Thus, we are currently working on a hierarchical model that is capable of modelling style.

\bibliography{paper}

\end{document}